# Spin injection into multilayer graphene from highly spin-polarized $Co_2FeSi$ Heusler alloy

Takehiro Yamaguchi[1], Rai Moriya[1,a)], Soichiro Oki[2], Shinya Yamada[2], Satoru Masubuchi[1], Kohei Hamaya[2], and Tomoki Machida[1,3,b)]

[1] *Institute of Industrial Science, University of Tokyo, 4-6-1 Komaba, Meguro, Tokyo 153-8505, Japan*

[2] *Graduate School of Engineering Science, Osaka University, 1-3 Machikaneyama, Toyonaka 560-8531, Japan*

[3] *Institute for Nano Quantum Information Electronics, University of Tokyo, 4-6-1 Komaba, Meguro, Tokyo 153-8505, Japan*

We demonstrate electrical spin injection into multilayer graphene (MLG) in a lateral spin valve device from a highly spin-polarized $Co_2FeSi$ (CFS) Huesler electrode. Exfoliated MLG was transferred onto pre-patterned epitaxial CFS wires grown on an Si(111) substrate by a polymer-based transfer method. This method enabled us to fabricate multiple single-crystal CFS electrodes in contact with MLG. Electrical spin injection from CFS to MLG was detected through non-local magnetoresistance (MR) measurement. A non-local spin signal of 430 Ω was observed; this is the largest value among all reported non-local MR values in graphene-based devices.

a) E-mail: moriyar@iis.u-tokyo.ac.jp
b) E-mail: tmachida@iis.u-tokyo.ac.jp

Two-dimensional (2D) crystals such as graphene, hexagonal boron nitride (h-BN), and transition metal dichalcogenides are emerging as a new material system and receiving considerable interest for both fundamental science and engineering points of view [1]. In bulk crystal, these materials have a structure consisting of 2D sheets connected vertically through the van der Waals (vdW) force. Because vdW interlayer coupling does not involve chemical bonding, these crystals can easily be exfoliated down to individual monolayers, or different crystals can be connected with the vdW force to form vdW heterostructure [2]. These vdW heterostructures have demonstrated potential for high-performance electronics and optoelectronics [3-5]. In particular, the long intrinsic spin relaxation time of graphene and graphene-based vdW heterostructures make them fascinating candidates for spintronic applications [6]. Therefore, significant efforts have been made to increase the spin relaxation time [7-13]. However, the spin injection efficiency need to be further improved to obtain a larger spin signal for use in spin-based device applications. Previous studies have shown that the spin injection efficiency can be improved by inserting a tunnel barrier between graphene and a ferromagnet [14]. However, the maximum spin injection efficiency is limited by the lack of a method to fabricate a high-quality tunnel barrier at the graphene/ferromagnet interface. As an alternative method, highly spin-polarized ferromagnetic materials can be used to achieve high spin injection efficiency without the use of a tunnel barrier [15,16]. However, spin injection from such highly spin-polarized material to graphene has not yet been demonstrated because most of these materials need to be a single crystal; this is very difficult to achieve on a graphene surface because of lattice mismatch and interface reaction. To overcome this issue, here we demonstrate that a lateral spin valve can be fabricated by transferring mechanically exfoliated multilayer graphene (MLG) onto a pre-patterned epitaxial Co$_2$FeSi (CFS) Heusler alloy. This is in contrast to the conventional fabrication of a lateral spin valve device, where the electrode is evaporated on the graphene surface [6]. Our presented inverted lateral spin valve structure enable the fabrication of multiple single-crystal CFS electrodes in electrical contact with the MLG. Our device demonstrated a large non-local magnetoresistance signal of 430 Ω with a moderate ferromagnetic contact separation of 4 μm. This is the largest non-local spin signal obtained so far in graphene-based devices, which we believe can be attributed to the electrical spin injection from the highly spin-polarized CFS electrode.

Fig. 1(a) schematically illustrates the device fabrication. By using molecular beam epitaxy, a highly ordered L2$_1$ structure CFS film with a thickness of 25 nm was epitaxially grown on a non-doped Si(111) substrate with a resistivity of 5 kΩ-cm [17]. Note that this is the only method for fabricating a high-quality single-crystal CFS film at the moment. Electron beam (EB) lithography and Ar$^+$ ion milling were used to pattern the CFS film into a wire shape with a contact pad at the end. The surface of the etched Si(111) substrate naturally oxidized once exposed to air after the ion milling. Separately, MLG with a thickness of ~3 nm was mechanically exfoliated from the Kish graphite and deposited onto a polymethyl methacrylate (PMMA)/glass slide. The dry transfer method [18] was used to transfer the MLG/PMMA stack onto the CFS wire at a substrate temperature of 120 °C; PMMA was subsequently dissolved in acetone. A similar method has recently been reported to fabricate an h-BN/graphene channel transferred onto a patterned Co/MgO electrode [12]. Here, we applied this method to fabricating the MLG/CFS junction. To fabricate the transparent contact at the MLG/CFS junction, a few volts of voltage with the duration of 1 second was applied under current limit of 1 μA between the CFS electrodes after the MLG was transferred onto the CFS wires at a temperature of 1.6 K. This induced local annealing of the MLG/CFS contact and reduced the contact resistance. Fig. 1(b) depicts the fabricated device structure. A lateral spin valve device composed of an MLG channel and multiple CFS contacts was fabricated. Because CFS is grown directly on the Si substrate, there is significant parallel conduction in the substrate at room temperature, which makes observation of non-local spin signals very difficult. To eliminate parallel conduction in the Si substrate, all of the measurements were performed at a low temperature (1.6 K) within a variable-temperature cryostat.

Figs. 2(a) and 2(b) show a top view and bird's-eye view, respectively, of a scanning electron microscopy (SEM) image taken of the fabricated MLG/CFS lateral spin valve device. To measure the magnetotransport, we made CFS electrodes with different width to produce different coercive fields with the application of the in-plane external magnetic field $B$. As shown in Fig. 2(b), the transferred MLG adhered on both the CFS wires and Si substrate; it was not suspended between CFS wires. Fig. 2(c) shows the two-terminal $I$–$V$ curve measured between contacts b and c at 1.6 K. The $I$–$V$ curve shows small nonlinearity around the zero bias and exhibits linearity at a high bias. Based on this, we think that the junction contained only a very thin tunnel barrier at the MLG/CFS interface and behaved similarly to an ohmic contact. The thin tunnel barrier is due to the native oxide present on the surface of the CFS film, which is generated during the device fabrication process. Considering the fact that the zero-bias resistance of 22 kΩ was obtained from the $I$–$V$ curve and the four-terminal resistance of MLG $R_{4t}$ ~ 1.2 kΩ was measured by applying a current between contacts a and d and measuring the voltage at contacts b and c, the contact resistance between the MLG and CFS was determined to be ~10 kΩ. This resistance was the same order as that of the reference sample, where MLG was transferred onto Pd wire using same transfer procedure; this sample exhibited a contact resistance of ~5 kΩ (see supplementary information). The contact resistance values of both CFS and Pd were higher than the typical contact resistance of a metal/graphene interface (less than 1 kΩ for a similar contact area) [19,20]. The nearly linear *I–V* curve with a large contact resistance suggests that the MLG/CFS junction fabricated by this method was a pinhole-like contact where the junction had a smaller effective area than actual contact size. This may have been due to both the finite roughness of the CFS wire surface and voltage application to the contact during fabrication; these points need to be improved upon in future experiments. We also fabricated a CFS wire for four-terminal measurement to evaluate its resistivity on the same wafer. The bulk spin polarization of CFS has a strong relationship with its resistivity [21]. The obtained resistivity of 56 μΩ-cm corresponded to the bulk spin polarization of the CFS wire used in this experiment of $P_F = 65\%–70\%$.

The non-local MR was measured at 1.6 K; a current was applied at contacts c and d, and the voltage was measured between contacts a and b under the application of the in-plane magnetic field $B_{ext}$. Fig. 3 shows the results, where the background MR signal has been subtracted from the measurement data. The non-local MR enabled us to eliminate all charge current-induced effects such as the anisotropic magnetoresistance of the CFS wire and to detect the signal solely due to the spin current-induced response. The clear non-local MR signal with hysteresis in Fig. 3 provides evidence of electrical spin injection into the MLG from the CFS electrode. The two resistance jumps at the low field and $B = 0.05$ T were due to the magnetization reversal of the wide and narrow CFS wires, respectively. The non-local MR was lower when two magnetic moments were anti-parallel; this is also consistent with the expectation from non-local MR with a symmetric spin injector and detector. When a current of 20 nA was applied for the measurement, the amplitude of the non-local MR was determined to be $\Delta R_{NL} = 430$ Ω. This is a significantly larger $\Delta R_{NL}$ value than those reported for graphene- and MLG-based spin valve devices [6]. Previously, $\Delta R_{NL} = 130$ Ω was obtained for a graphene spin valve device using a Co/MgO/TiO$_2$ contact with an electrode separation of 2.1 μm; the spin injection efficiency was estimated to be $P_J = 26\%$–$30\%$ [14]. We obtained larger $\Delta R_{NL}$ with a greater electrode separation of 4 μm in the MLG device with a CFS contact. These comparisons suggest that the increased $\Delta R_{NL}$ is related to the use of the highly spin-polarized CFS electrode, which may increase the spin injection efficiency compared to conventional ferromagnetic materials.

To estimate the spin injection efficiency $P_J$ in our device, we followed the spin-drift diffusion model proposed by Takahashi *et al.* [22] and obtained $\Delta R_{NL} \cong P_J^2 R_G e^{-L/\lambda_G}$, where $R_G = R_{4t}\lambda_G/L$ denotes the spin resistance of MLG, $R_{4t} = 1.2$ kΩ denotes the four-terminal resistance of MLG, $\lambda_G$ is the spin diffusion length of MLG, and $L = 4$ μm is the distance between two inner CFS electrodes. This approximated form is valid because the contact resistance at the MLG/CFS interface $R_J \sim 10$ kΩ was much larger than $R_G$ for typically observed $\lambda_G$ values for graphene and MLG. Fig. 4 plots the relation between $P_J$ and $\lambda_G$ in our device calculated for $\Delta R_{NL} = 430$ Ω. This figure can be used to determine the spin injection efficiency if $\lambda_G$ is given. Because it is difficult to determine the spin diffusion length from a Hanle measurement [6] on our device, due to the small out-of-plane anisotropy field of the CFS film, we speculate on the value of $\lambda_G$ from literature. Reported $\lambda_G$ values for graphene or MLG fabricated on an $SiO_2$ substrate have ranged from 1 to 8 μm [8,10,14,23-28]. Even if we assumed the longest reported spin diffusion length of 8 μm for MLG, the spin injection efficiency $P_J$ can be as high as 55%. Moreover, the spin injection efficiency $P_J$ should not exceed $P_F$ unless we use a single crystalline tunnel barrier such as MgO in between CFS and MLG. Thus, the $P_F$ value provides the maximum limit of $P_J$ expected to achieve in our device and this value is indicated with blue dotted line in Fig. 4. Therefore, the possible range of $P_J$ in our device is determined as $P_J$= 55%−70%. The large $P_J$ value is clear evidence that the large $\Delta R_{NL}$ we obtained was due to the improved spin injection efficiency with the use of the epitaxial CFS compound. A more precise determination of the spin diffusion length from the Hanle measurement or measurements on different CFS electrode separations could be done in future experiments with further optimization of the device fabrication method.

In summary, we demonstrated the dry transfer fabrication of a lateral MLG spin valve device with a CFS electrode. A large non-local MR signal of 430 Ω was demonstrated at 1.6 K. The significantly large non-local MR compared with previously reported values was due to the highly spin-polarized nature of the CFS compound and demonstrates the potential applicability of this material to graphene-based spintronic devices.

**Acknowledgements**

This work was partly supported by a Grant-in-Aid for Scientific Research on Innovative Areas "Science of Atomic Layers" and "Nano Spin Conversion Science" from the Ministry of Education, Culture, Sports, Science and Technology (MEXT) and the Project for Developing Innovation Systems of MEXT and Grants-in-Aid for Scientific Research from the Japan Society for the Promotion of Science (JSPS) and CREST, Japan Science and Technology Agency (JST).

**Figure captions**

Figure 1

(a) Schematic illustration of the lateral spin valve device fabrication using dry transfer of an MLG flake. The MLG/PMMA/glass plate structure was fabricated by mechanical exfoliation and deposition of MLG. Separately, patterned epitaxial CFS wire was patterned on Si(111). Under the microscope observation, these two structures were aligned and placed into contact. (b) Illustration of the lateral spin valve structure. The current source and voltmeter were connected. Arrows depict the direction of magnetization of the CFS electrode.

Figure 2

(a) Top view and (b) bird's-eye view of scanning electron microscope (SEM) images of the MLG/CFS lateral spin valve device. The direction of the external magnetic field $B$ is indicated by the arrow. (c) Current–voltage ($I$–$V$) characteristics measured between contacts b and c in (a) at 1.6 K.

Figure 3

(a) Non-local magnetoresistance data measured at 1.6 K. A current was applied between contacts c and d to inject spin-polarized electrons. The spin current was detected by measuring the voltage between contacts a and b. The direction of magnetization of contacts b and c are indicated by arrows.

Figure 4

Relation between the spin injection efficiency $P_J$ and spin diffusion length of MLG $\lambda_G$ derived using parameter for the device shown in Fig. 2. This relation can be used to determine the spin injection efficiency $P_J$ if $\lambda_G$ is given. The blue dotted line indicates the maximum $P_F$ value for the CFS electrode.

Figure 1

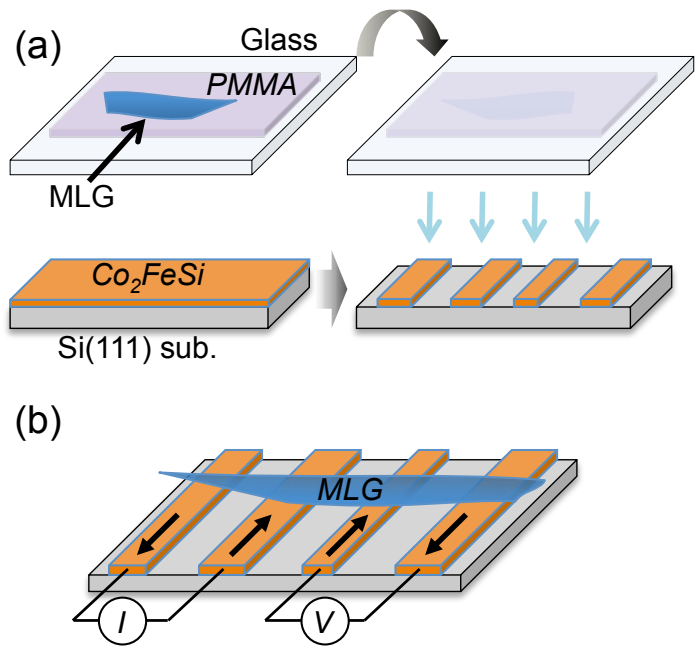

Figure 2

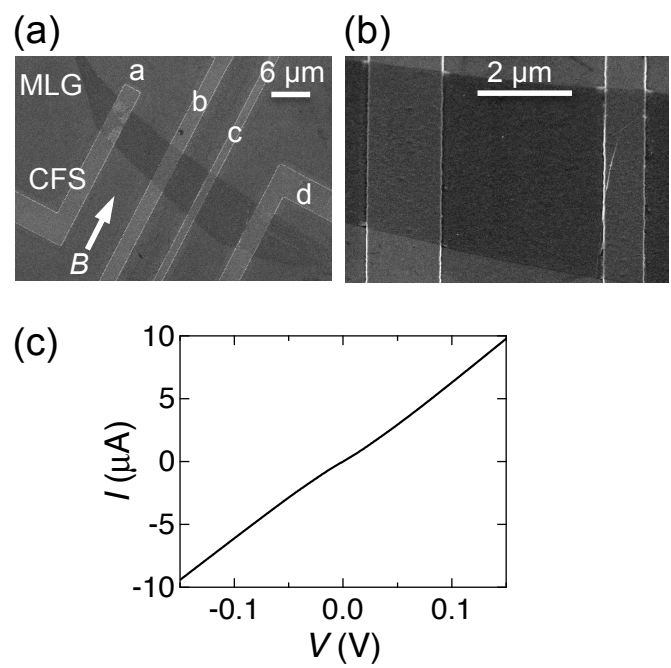

Figure 3

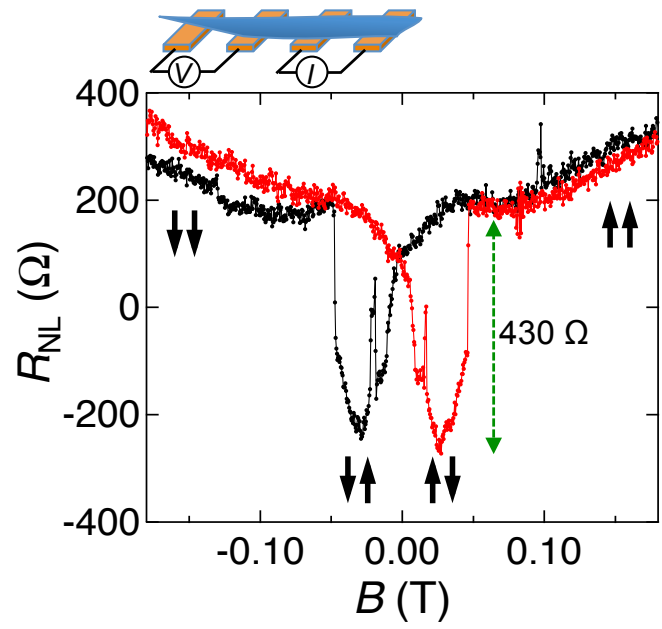

Figure 4

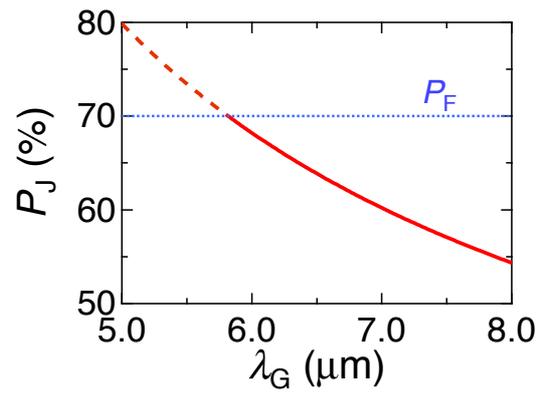